\documentclass[aps,pre,showpacs,superscriptaddress]{revtex4}
\usepackage{graphicx}
\usepackage{color}
\usepackage{amsmath}

\begin{document}

\title{Stability and collapse of localized solutions of the controlled 
three-dimensional Gross-Pitaevskii equation}

\author{Renato Fedele}
\email{renato.fedele@na.infn.it} \affiliation{Dipartimento di
Scienze Fisiche, Universit\`{a} Federico II and INFN Sezione di
Napoli, Complesso Universitario di M.S. Angelo, via Cintia,
I-80126 Napoli, Italy, EU}
\author{Dusan Jovanovi\'c}
\email{djovanov@phy.bg.ac.yu} \affiliation{Institute of Physics,
P. O. Box 57, 11001 Belgrade, Serbia}
\author{Bengt Eliasson}
\email{bengt@tp4.rub.de} \affiliation{Institut f\"ur
Theoretische Physik IV, Ruhr--Universit\"at Bochum, D-44780
Bochum, Germany, EU}
\affiliation{\small Department of Physics, Ume{\aa} University, SE-90 187 Ume{\aa}, Sweden}
\author{Sergio De Nicola}
\email{s.denicola@cib.na.cnr.it} \affiliation{Istituto di
Cibernetica ``Eduardo Caianiello'' del CNR Comprensorio ``A.
Olivetti'' Fabbr. 70, Via Campi Flegrei, 34, I-80078 Pozzuoli
(NA), Italy, EU} \affiliation{Dipartimento di Scienze Fisiche,
Universit\`{a} Federico II and INFN Sezione di Napoli, Complesso
Universitario di M.S. Angelo, via Cintia, I-80126 Napoli, Italy, EU}
\author{Padma Kant Shukla}
\email{ps@tp4.rub.de} \affiliation{Institut f\"ur Theoretische
Physik IV, Ruhr--Universit\"at Bochum, D-44780 Bochum, Germany, EU}
\date{\today}
{\small\begin{abstract} On the basis of recent investigations, a newly developed analytical procedure is used for constructing
a wide class of localized solutions of the controlled three-dimensional
(3D) Gross-Pitaevskii equation (GPE) that governs the dynamics of
Bose-Einstein condensates (BECs). The controlled 3D GPE is decomposed into a two-dimensional (2D) linear Schr\"{o}dinger equation and a one-dimensional (1D) nonlinear Schr\"{o}dinger equation, constrained by a variational condition for
the controlling potential. Then, the above class of localized solutions are constructed as the product of the solutions of the transverse and longitudinal equations. On the basis of these exact 3D analytical solutions, a stability analysis is carried out, focusing our attention on the physical conditions for having collapsing or non-collapsing solutions.
\end{abstract}}

\maketitle

\section{Introduction}\label{Introductory}
About 85 years ago, the seminal work of Bose \cite{Bose} opened up the study 
of the statistical properties of bosons in ultra-cold quantum systems. 
Bose's idea was further developed by Einstein \cite{Einstein}, leading to the 
theoretical prediction of the condensation of atoms in the lowest quantum state 
below a certain temperature. The idea of Bose-Einstein of atom condensation in 
the ground state has been experimentally verified in a dilute gas composed of 
atoms \cite{Ketterle,Anderson,Bradley,Davis}. The dynamics of nonlinearly 
interacting bosons in ultra-low temperature gases is governed by
the Gross-Pitaevskii equation (GPE), which is an extension of the nonlinear 
Schr\"{o}dinger equation (NLSE) by including the confining potential and
inter-atomic interactions \cite{Gross-Pitaevskii}. The GPE, without the 
external potential, admits localized solutions in the form of one-dimensional 
dark and bright solitons, as well as radially symmetric vortex structures. 
Nonlinear localized excitations involving BECs arise due to the balance between 
the spatial dispersion of matter waves and nonlinearities caused by 
repulsive or attractive inter-atomic interactions in BECs.
Recent observations  \cite{r1a,r1b,r1c,r1d,r2a,r2b,r2c,r2d} conclusively 
demonstrated the existence of bright \cite{r2a,r2b,r2c,r2d} and dark/grey 
\cite{r1a,r1b,r1c,r1d} matter wave solitons and quantum vortices \cite{quantum-vortices}.

Although the area of investigations of localized solutions of the
GPE is quite fascinating, most of the theoretical results deals
with approximate solutions in 3D or in reduced geometries \cite{5}. 
They are well supported by suitable numerical evaluations \cite{6} and
adequately compared with a very broad spectrum of experimental
observations (for a review, see Ref.s \cite{JPhysB,Kevrekidis-et-al}.) 
Nevertheless, this testifies that finding exact localized solutions 
of the 3D GPE in a trapping external potential well, and preserving 
their stability for a long time, is still a challenging task. In
particular, one encounters serious difficulties in attempting to
find soliton solutions in one or more spatial dimensions, although
several kinds of solitons have been found using certain approximations
\cite{9,Carr02}. This leads us to arrive at the conclusion that, in order to have exact soliton solutions in BECs, some sort of "control of the
system" may be necessary.

The very large body of experience suggests that interacting bosons
constitute a nonlinear and nonautonomous system \cite{7}, for which coherent
stationary structures (i.e. solutions of the 3D GPE) exist only
if suitable time-dependent external potentials are taken "ad hoc"
\cite{8}. Therefore, the correct analysis of the system should
include a 'controlling potential' term in the GPE, to be determined
self-consistently with the desired solutions (e.g. the localized
solutions). This procedure may be, in principle, extended to an
arbitrary 'controlled solution' with the appropriate choice of the
controlling potential \cite{10}. The controlling potential method
(CPM) has been proposed in the literature, and used to find the
multi-dimensional controlled localized solutions of the GPE
\cite{10}. Preliminary investigations \cite{11} have suggested
that control operations introduced by this method ensures the
stability of coherent solutions against relatively small errors
in experimental realizations of the prescribed controlling
potential. This idea could be realized by techniques that
involve lithographically designed circuit patterns, providing the
electromagnetic guides and microtraps for ultracold systems of
atoms in BEC experiments \cite{Forthagh98}, and by the optically
induced 'exotic' potentials \cite{Grimm00}.

Another important aspect of solitons in BECs is the phenomenon of 
collective collapse/explosion, that has been predicted \cite{Carr02} 
and observed experimentally \cite{Sackett99,Donley01}. In particular, this 
phenomenon seems to be dependent on the parameters of the BECs and
on the confining or repulsive potential \cite{Carr02}. 

The stabilization and control of BECs in asymmetric traps have
been investigated via time-dependent solutions of the GPE
\cite{Garcia98}. Stable condensates, with the limited number of
$^7$Li atoms with attractive interaction, have been observed in a
magnetically trapped gas \cite{Bradley97}.

Recently, a mathematical investigation oriented towards the
construction of 3D analytical solutions of the controlled GPE has
been carried out \cite{GPE1} and applied to the construction 
of 3D exact localized solutions \cite{pre2009}. 
In Ref. \cite{GPE1}, it has been proven that, under the
assumption of the separability of the external trapping potential
well $V_{trap}$ in the spatial coordinates [viz. $V_{trap}(x,y,z,t)
= V_\bot (x,y,t) + V_z (z,t)$, where $V_\bot$ and $V_z$ are referred
to as the 'transverse' and  the 'longitudinal' potentials,
respectively], and a suitable constrained variational condition for the
controlling potential $V_{contr}$ (i.e. the average over the
'transverse' $x-y$ plane of $V_{contr}$ is required to be a
stationary functional of the BEC's transverse profile), the
factorized form of the solution of the 3D controlled GPE, in the
form $\psi (x,y,z,t) = \psi_\bot (x,y,t) \psi_z (z,t)$, can be
constructed, so that $\psi_\bot$ and $\psi_z$ satisfy a 2D linear
Schr\"{o}dinger equation and a 1D nonlinear Schr\"{o}dinger
equation, respectively.

In this paper, we apply the results of the above investigation \cite{GPE1} to
develop a new analytical procedure for constructing a broad class
of exact localized solutions of the controlled 3D GPE, with a
parabolic external potential well. In particular, we extend our recent 
investigation \cite{pre2009} to a wider family of exact 3D localized solutions
of the controlled GPE and perform an analysis that establishes 
the physical conditions and parameter regimes for having collapsing and 
non-collapsing localized solutions.
In the next section, we formulate
our problem and present the controlled GPE, and we briefly
summarize the results found in Ref. \cite{GPE1}. In section
\ref{exact-solutions}, we apply these results to obtain localized
solutions of the controlled GPE in the form of bright, dark and grey
solitons for the longitudinal profile $|\psi_z|^2$, and in
the form of the Hermite-Gauss functions for the transverse profile
$|\psi_\bot|^2$. We use the decomposition procedure of the
controlled GPE suggested in Ref.\,\cite{GPE1} and solve the 2D transverse
linear Schr\"{o}dinger equation to obtain $\psi_\bot$ in terms of
Hermite-Gauss modes. Then, the 1D longitudinal controlled NLSE is solved using a method based on the Madelung's fluid 
representation \cite{Fedele02,FedeleSchamel02}, which separates the 
NLSE into a pair of equations, composed of one continuity equation and one Korteweg-de Vries (KdV)-type equation. It is shown that the phase of the 
longitudinal wavefunction $\psi_z$ has a parabolic dependence on the
variable $z$. As the transverse and longitudinal equations are
coupled both through the coefficient of the nonlinear term (in the
longitudinal equation) and through the controlling potential, the
consistency condition between the transverse and longitudinal
solutions set up a relationship between the transverse and
longitudinal restoring forces of the external trapping potential
well. From the latter, the explicit spatio-temporal dependence of
the controlling potential is self-consistently determined in each
particular case. In section \ref{collapse} a detailed analysis of 
the dynamics of the above exact 3D solutions is developed both 
analytically and numerically. In particular, we study 
the properties of the controlled BEC states in the 2D case, showing  
that they feature breathing (oscillations of the amplitude and 
position) due to the oscillations of the perpendicular solution, 
as well as the oscillations in the parallel direction, arising from the 
initial displacement of the structure from the bottom of the potential well
in the parallel direction. A stability analysis of controlled GPE 
structures is carried out in terms of a set of six parameters related
through four equations. Therefore, two of them can be assumed arbitrarily. 
We discuss some examples of parameters corresponding to collapsing 
and non-collapsing 3D solutions. Finally, the conclusions are summarized in section \ref{conclusion}.

\section{Decomposition of the Controlled Gross-Pitaevskii equation}
\label{controlled-GPE}

It is well known that the spatio-temporal evolution of an ultracold
system of identical atoms forming a Bose Einstein condensate (BEC)
in the presence of an external potential $U_{ext}(\mathbf{r},t)$,
within the mean field approximation, is governed by the 3D GPE,
viz.
\begin{equation}\label{GPE-0}
    i\mathop \hbar\frac{\partial \Psi }{\partial
t}=-\frac{\mathop \hbar }{2m_a}\nabla ^2\Psi + N Q\left| \Psi
\right|^2\Psi + U_{ext} \left(\mathbf{r},t\right)\,\Psi\,,
\end{equation}
where $\Psi (\mathbf{r},t)$ is the wavefunction describing the BEC
state, $m_a$ is the atom mass, $Q$ is a coupling coefficient
related to the short range scattering (s-wave) length $a$
representing the interactions between atomic particles, i.e.
$Q=4\pi\hbar^2a/m_a$, and N is the number of atoms. Note that the
short range scattering length can be either positive or negative.
Solitons have been observed in BECs of $^7$Li atoms with a small
scattering length $a \approx -0.2\,\mathrm{nm}$, in correspondence of
the following typical values of the parameters: $N=10^4$--$10^5$  at 
a temperature of $1$--$10\,{\rm \mu K}$ and a magnetic field $\sim 400$--$600\,\mathrm{G}$  \cite{r1a,r2b}.

We assume that $U_{ext}$ is the sum of a 3D trapping potential
well, say $V_{trap}$, to confine the particles of a BEC, and a
controlling potential, say $V_{contr}$, to be determined
self-consistently. Furthermore, under this assumptions, we
introduce the variable $s=ct$ ($c$ being the speed of light) and
divide both sides of Eq.(\ref{GPE-0}) by $m_a c^2$, in such a way
that
\begin{equation}\label{ext-potential}
  \frac{U_{ext}(\mathbf{r}, t)}{m_a c^2} \equiv V_{trap}(\mathbf{r},s) +
  V_{contr}(\mathbf{r},s)\,,
\end{equation}
and Eq. (\ref{GPE-0}) can be cast into the form
\begin{equation}\label{3D-GPE}
    i\mathop \lambda \nolimits_c \frac{\partial \psi }{\partial
s}=-\frac{\mathop \lambda \nolimits_c^2 }{2}\nabla ^2\psi + \left[
V_{trap}(\mathbf{r},s) + V_{contr} \left(\mathbf{r},s\right) + q
|\psi|^2\right]\,\psi\,,
\end{equation}
where $\psi(\mathbf{r},s) \equiv \Psi (\mathbf{r},t=s/c)$,
$\mathop \lambda \nolimits_c \equiv \hbar/m_a c^2$ is the Compton
wavelength of the single atom of the BEC and $q \equiv N Q/mc^2$.

In order to decompose Eq. (\ref{3D-GPE}), we briefly summarize 
the results of Ref.\,\cite{GPE1}. To this end, we first assume that
\begin{equation}\label{V-trap}
    V_{trap}(\mathbf{r},s) = V_\bot (\mathbf{r}_\bot,s) + V_z (z,s)\,,
\end{equation}
where, in Cartesian coordinates, $\mathbf{r} \equiv (x,y,z)$ and
$\mathbf{r}_\bot \equiv (x,y)$ denotes, by definition, the
'transverse' part of the particle's vector position $\mathbf{r}$ and
$z$  the 'longitudinal' coordinate. Additionally, let us denote, 
in Cartesian coordinates, $\nabla_\bot \equiv
\hat{x}\,\partial/\partial x + \hat{y}\,\partial/\partial y$. 

Let us suppose that $\psi_\bot (\mathbf{r}_\bot,s)$ and $\psi_z (z,s)$ are two complex functions satisfying the following  2D linear Schr\"{o}dinger equation
\begin{equation}\label{2D-LSE}
  \left(i\mathop \lambda \nolimits_c \frac{\partial}{\partial
s} -\widehat{H}_\bot\right)\psi_\bot =0\,,
\end{equation}
where
\begin{equation}\label{H-perp}
    \widehat{H}_\bot = - \frac{\mathop \lambda \nolimits_c^2 }{2}\nabla_\bot^2 + V_\bot (\mathbf{r}_\bot,s)
\end{equation}
and the following 1D NLSE
\begin{equation}\label{1D-NLSE}
    \left(i\mathop \lambda \nolimits_c \frac{\partial}{\partial
s} -\widehat{H}_z \right)\psi _z = 0\,,
\end{equation}
where
\begin{equation}\label{H-z}
    \widehat{H}_z = - \frac{\mathop \lambda \nolimits_c^2 }{2}\frac{\partial^2}{\partial z^2} + V_z (z,s) +
    q_{1D}(s)\,|\psi_z (z,s)|^2 + V_0\,,
\end{equation}
respectively. In the latter, $V_0$ is an arbitrary real constant and the function $q_{1D}(s)$ 
is defined as 
\begin{equation}\label{q1D-definition}
    q_{1D}(s) = q\,\int d^2\vec r_\bot\,
    \left|\psi_\bot\right|^4\,.
\end{equation} 
Hereafter $q_{1D}$ is referred to as the 'controlling parameter'.

Furthermore, let us assume that the controlling potential depends in principle on $\mathbf{r}_\bot$ 
through $|\psi_\bot|^2$, viz. $V_{contr} = V_{contr}
\left(\rho_\bot (\mathbf{r}_\bot,s),z,s\right)$,
where $\rho_\bot (\mathbf{r}_\bot,s) \equiv |\psi_\bot
(\mathbf{r}_\bot,s)|^2$. 

Provided that
\begin{equation}\label{V-contr-stationary}
    V_{contr}(\mathbf{r}_\bot,z,s) = \left[q_{1D}(s) - q
    |\psi_\bot (\mathbf{r}_\bot,s)|^2\right]|\psi_z (z,s)|^2 + V_0\,,
\end{equation}
which makes stationary the functional (with respect to variation $\delta\rho_\bot$ of $\rho_\bot$; $z$ and $s$ play here a role of parameters)
\begin{equation}\label{functional-V-cal}
    {\cal V}[\rho_\bot; z,s]= \int
    \rho_\bot(\mathbf{r}_\bot,s)\,V_{contr}
    \left(\rho_\bot
    (\mathbf{r}_\bot,s),z,s\right)\,d^2\mathbf{r}_\bot\,,
\end{equation} 
under suitable constraints provided by the normalization condition for $\psi_\bot$, viz. $\int \rho_\bot (\mathbf{r}_\bot,s)\,d\mathbf{r}_\bot =1$, and by Eq. (\ref{q1D-definition}), where $q_{1D}$ is thought as a given function of $s£$, it can been shown \cite{GPE1} that the complex function
\begin{equation}\label{psi-product}
    \psi (\mathbf{r},s) = \psi_\bot (\mathbf{r}_\bot,s)\, \psi_z (z,s)
\end{equation}
is a 3D solution of the controlled Gross-Pitaevskii equation (\ref{3D-GPE}).

In summary, according to the results given in Ref.\,\cite{GPE1}, one may construct 3D solutions in the factorized form (\ref{psi-product}) such that the 3D controlled GPE is decomposed into the set of coupled equations (\ref{2D-LSE})
and (\ref{1D-NLSE}) plus the self-consistent expression of the controlling potential  (\ref{V-contr-stationary}) coming from a constrained variational condition.
Once Eqs. (\ref{2D-LSE}) and (\ref{1D-NLSE}) are solved, 
$\psi_\bot (\mathbf{r}_\bot,s)$ and $\psi_z (z,s)$ become
known functions. Consequently, the explicit space and time
dependence of $V_{contr}$ is self-consistently determined, as well, i.e.
the appropriate controlling potential corresponding to the
controlled solution (\ref{psi-product}) is deduced.

Note that, since $\psi_\bot $ satisfies Eq. (\ref{2D-LSE}),
according to the definition (\ref{functional-V-cal}), ${\cal V}$
represents the average of $V_{contr}$ in the transverse plane. The
value of this average corresponds to the arbitrary constant $V_0$.
Without loss of generality, we may put $V_0 = 0$, viz.
\begin{equation}\label{mimimize_2-D}
  \int d^2\vec r_\bot \,\, \psi_\bot^*\, V_{contr}\,\psi_\bot =0 .
\end{equation}
In this way, among all possible choices of $V_{contr}$, we adopt the
one which does not change the mean energy of the system (the
average of the Hamiltonian operator in Eq. (\ref{3D-GPE}) is the
same with or without $V_{contr}$) and therefore minimizes the
effects introduced by our control operation.

In the next section, we will apply the results obtained here to
the case of parabolic potentials, $V_\bot$ and $V_z$, to give
exact 3D controlled localized solutions of Eq. (\ref{3D-GPE}).

\section{Exact localized solutions of the controlled 3D GPE with a 3D parabolic potential well}\label{exact-solutions}

Let us assume that $V_\bot (\mathbf{r}_\bot,s)$ and $V_z (z,s)$ are
the usual parabolic potential wells to confine the particle of a
BEC, viz.
\begin{equation}\label{V-perpendicular}
  V_\bot (\mathbf{r}_\bot, s) = \frac{1}{2}\left[ {\omega_x^2\left(s\right)
x^2+\omega_y^2\left(s\right) y^2} \right] \,,
\end{equation}
and
\begin{equation}\label{V-zlongitudinal}
  V_z (z, s) = \frac{1}{2}\,\omega_z^2\left(s\right) z^2
\,,
\end{equation}
where, in general, the frequency $\omega_j$, $j=x,y,z$, are
supposed functions of time. The standard confining potential wells
(along each direction) corresponds to the assumption that they are
real quantity (positivity of their squares). However, our analysis
can be extended to the case in which they are assumed imaginary
(negativity of their squares).

It follows that Eq. (\ref{3D-GPE}) becomes
\begin{equation}\label{GPE}
    i\mathop \lambda \nolimits_c \frac{\partial \psi }{\partial
s}=-\frac{\mathop \lambda \nolimits_c^2 }{2}\nabla ^2\psi
+\frac{1}{2}\left[ {\omega_x^2\left(s\right)
x^2+\omega_y^2\left(s\right) y^2+\omega_z^2\left(s\right) z^2}
\right]\psi +q\left| \psi \right|^2\psi + V_{contr}
\left(x,y,z,s\right)\,\psi\,.
\end{equation}
According to the results and the assumptions of the previous
sections, if we seek a solution in the factorized form
\begin{equation}\label{101}
    \psi(x,y,z,s) = \psi_\bot(x,y,s)\,\, \psi_z(z,s),
\end{equation}
Eq. (\ref{GPE}) can be decomposed into the following set of
equations:
\begin{equation}\label{linsch}
    i\mathop \lambda \nolimits_c \frac{\partial \psi _\bot }{\partial
s}+\frac{\mathop \lambda \nolimits_c^2 }{2}\nabla _\bot ^2\psi
_\bot -\frac{1}{2}\left[ {\omega_x^2\left(s\right)
x^2+\omega_y^2\left(s\right) y^2} \right]\psi _\bot =0,
\end{equation}
\begin{equation}\label{104}
    i\mathop \lambda \nolimits_c \frac{\partial \psi _z\left(z,s\right) }{\partial
s}+\frac{\mathop \lambda \nolimits_c^2 }{2} \frac{\partial ^2 \psi
_z\left(z,s\right)}{\partial z^2}
 -\frac{1}{2}\,\omega_z^2\left(s\right) z^2\psi _z \left(z,s\right) - q_{1D} (s)\left| {\psi _z \left(z,s\right)}
\right|^2\psi _z \left(z,s\right) =0,
\end{equation}
\begin{equation}\label{controlling-potential}
    V_{contr}(x,y,z,s) = \left[q_{1D}(s) - q
    |\psi_\bot (\mathbf{r}_\bot,s)|^2\right]|\psi_z (z,s)|^2 \,.
\end{equation}

\subsection{Solution of the transverse equation with a 2D parabolic
potential well}\label{Perpendicular solution} 
Equation (\ref{linsch}) is readily solved, and its solution can be found in the standard
literature, but we present it here for completeness. The general
solution $\psi_\bot(x,y,t)$ can be expressed as the superposition
of Hermite-Gauss modes, viz.
\begin{equation}\label{superposition}
\psi_\bot\left(x,y,s\right) =   \sum _{n=0}^{\infty } \sum
_{l=0}^{\infty } \alpha_{n,l}\,
   \psi_{x,n}\left(x,s\right) \, \psi_{y,l}\left(y,s\right)
\end{equation}
where $\alpha_{n,l}$ are arbitrary constants and
\begin{equation}\label{HermiteGauss}
\psi_{j,k}\left(j,s\right) =
   \left[\pi\, 2^{2 k+1} (k!)^2 \sigma_j^2\left(s\right)\right]^{-\frac{1}{4}}\,
   \exp\left[\frac{i \gamma_j\left(s\right) j^2}{2 \lambda_c}-\frac{j^2}{4\,\sigma_j^2\left(s\right)} +
   i \phi_{j,k}\left(s\right)\right]\,
   H_k\left[\frac{j} {\sqrt{2}\, \sigma_j
\left(s\right)}\right]\,,
\end{equation}
where $j=x,y$. The perpendicular spatial scale $\sigma_j$ (i. e.,
the root of mean square) of the Hermite-Gauss functions
(\ref{HermiteGauss}) satisfies the Ermakov-Pinney equation \cite{Ermakov,Pinney}
\begin{equation}\label{Pinnerm}
\frac{d^2\sigma_j\left(s\right)}{ds^2} + \omega_j^2(s)
 \sigma_j\left(s\right) -\frac{\lambda_c^2
}{4\,\sigma_j^3\left(s\right) } = 0,
\end{equation}
and the phase functions $\gamma_j (s)$ and $\phi_{j,k}(s)$ are
given by
\begin{equation}\label{gamma}
 \gamma_j\left(s\right) =
 \frac{\sigma_j'\left(s\right)}{\sigma_j\left(s\right)},
\end{equation}
\begin{equation}\label{phi}
 \phi_{j,k}\left(s\right) = \phi_{j,k,0}-\frac{\lambda_c }{4} \,(2 k+1)\int_0^s
 \frac{ds'}{\sigma_j^2\left(s'\right)},
\end{equation}
where $\phi_{j,k,0}$ is an arbitrary constant.

\subsection{Exact solution of the longitudinal NLSE with a
1D parabolic trapping potential well}\label{parallel_solution} 
In order to solve Eq. (\ref{104}), we first observe that, according to
Eqs. (\ref{q1D-definition}), (\ref{superposition}) and
(\ref{HermiteGauss}), the controlling parameter $q_{1D}$ can be
expressed in terms of the features of $\Psi_\bot$, viz.
\begin{equation}\label{q1D}
  q_{1D}(s) = q\,\frac{ {\cal F}\left[\sigma_x (s), \sigma_y
  (s)\right]}{\sigma_x\left(s\right)\sigma_y\left(s\right)}\,,
\end{equation}
where ${\cal F}\left[\sigma_x (s), \sigma_y(s), s\right]$ is a
(relatively complicated) positive definite functional of
$\sigma_x$ and $\sigma_y$ given by
{\small\begin{equation}\label{functional-F}
    {\cal F}\left[\sigma_x (s), \sigma_y(s)\right] = \int\,\frac{e^{-2u^2}du}{\sqrt{2}\,\pi}\,
    \int\,\frac{e^{-2v^2}dv}{\sqrt{2}\,\pi}\,\left|\sum_{n,m,l,p}{\cal F}_{n,m,l,p}\left[\sigma_x (s), \sigma_y(s)\right]H_n(u)H_m(u)H_l(v)H_p(v)\right|^2\,,
\end{equation}}
with $u=x/\sqrt{2}\sigma_x$, $v=y/\sqrt{2}\sigma_y$ and
{\small\begin{eqnarray}\label{F-nmlp}
{\cal F}_{n,m,l,p}\left[\sigma_x (s), \sigma_y(s)\right] &=&
\frac{\alpha_{nl}\alpha^{*}_{mp}}{\sqrt{2^{n+m+l+p}\,n!\,m!\,l!\,p!}}\,
e^{i\left(\phi_{x,n,0}-\phi_{x,m,0}\right)}\nonumber\\
&\times& e^{i\left(\phi_{y,l,0}-\phi_{y,p,0}\right)}\,e^{-\frac{i\mathop
\lambda \nolimits_c}{2}\left[(n-m)\int_0^s\frac{ds'}{\sigma^{2}_x
(s')}+ (l-p)\int_0^s\frac{ds''}{\sigma^{2}_y (s'')}\right]}
\end{eqnarray}}

From the above equation, it is clear that, in general, $q_{1D}$
depends in a non-trivial manner on $\sigma_x (s)$ and $\sigma_y (s)$.
However, in the simple case when the perpendicular solution contains
{\em only one} Gauss-Hermite mode, say the $(n,l)$-mode, Eqs.
(\ref{superposition}) and (\ref{HermiteGauss}), ${\cal F}$ becomes a real
positive constant, viz.,
\begin{equation}\label{functional-F-bis}
{\cal F}\left[\sigma_x (s), \sigma_y (s)\right] =
\int\,\frac{e^{-2u^2}\,H_n^4 (u)\,du}{\sqrt{2}\,\pi}\,
    \int\,\frac{e^{-2v^2}\,H_l^4 (v)\,dv}{\sqrt{2}\,\pi} \equiv \delta_n\,\delta_l = {\rm constant}
\end{equation}
and, therefore, $q_{1D}$ becomes
\begin{equation}\label{q1D-bis}
q_{1D}(s) = q\,\frac{
\delta_n\,\delta_l}{\sigma_x\left(s\right)\sigma_y\left(s\right)}\equiv
q_{1D}^{n,l}(s)\,.
\end{equation}

\subsubsection{Reduction of the 1D NLSE to a KdV-like equation by means of the Madelung's fluid
approach \label{Madelung}} 
Equation (\ref{104}) is a 1D GPE with a time dependent parabolic potential. Its
approximate solutions are well known in the literature, but we
attempt here to find an {\it exact} solution which is also
compatible with the ones of the transverse part, in such a way to
give a solution of the full controlled 3D GPE (\ref{GPE}).
We seek $\psi_z(z,s)$ using the following standard Madelung's
fluid representation
\begin{equation}\label{2}
\psi_z\left(z,s\right) = \sqrt{\rho (z,s)}\, \exp\left[{\frac{i \Theta
(z,s)}{\lambda_c }}\right],
\end{equation} which, after the substitution into Eq. (\ref{104})
and the separation of real and imaginary parts, yields
\begin{equation}
\label{H-J-eq} \frac{\partial\Theta}{\partial s} + \frac{1}{2}\,
\left(\frac{\partial\Theta}{\partial z}\right)^2 + U(z,s) -
\frac{\lambda_c^2}{2}\frac{1}{\rho^{1/2}}
\frac{\partial^2\rho^{1/2}}{\partial z^2}=0\,,
\end{equation}
   and
\begin{equation}\label{4}
\frac{\partial \rho}{\partial s} + \frac{\partial}{\partial
z}\left(\rho\frac{\partial\Theta}{\partial z}\right)
 = 0\,,
 \end{equation}
 where
\begin{equation}\label{longitudinal-potential}
U(z,s) = \frac{1}{2}\,\omega_z^2 (s)\,z^2 + q_{1D}(s)\,\rho (z,s)
\end{equation}
 is the 'longitudinal potential energy' which is a functional of $\rho$.

By differentiating Eq. (\ref{H-J-eq}) with respect to $z$ and
introducing the 'current velocity' $V\equiv
\partial\Theta/\partial z$, a non-trivial series of transformations
given in Ref.s \cite{Fedele02,FedeleSchamel02} allows us to obtain the following
generalized KdV equation
\begin{equation}
\label{motion-eq} -\rho\frac{\partial V}{\partial
s}+V\frac{\partial \rho}{\partial s}
 + 2\left[c_0(s)-\int^z\frac{\partial V}{\partial s}dz
 \right]\frac{\partial \rho}{\partial z}
-\left(\rho\frac{\partial U}{\partial z}+2U\frac{\partial
\rho}{\partial z}
 \right)+\frac{1}{4}\frac{\partial^3\rho}{\partial z^3}=0\,,
\end{equation}
where $c_0 (s)$ is an arbitrary function of $s$. Making use of
definition (\ref{longitudinal-potential}), Eq. (\ref{motion-eq})
becomes
\begin{equation}
\label{motion-eq-1} -\rho\frac{\partial V}{\partial
s}+V\frac{\partial \rho}{\partial s}
 + 2\left[c_0(s)-\int^z\frac{\partial V}{\partial s}dz
 \right]\frac{\partial \rho}{\partial z}
-\omega_z^2(s)\,z\rho
-\omega_z^2(s)\,z^2\frac{\partial\rho}{\partial z}-
3q_{1D}(s)\,\rho\frac{\partial\rho}{\partial z} +
\frac{\lambda_c^2}{4}\frac{\partial^3\rho}{\partial z^3}=0\,.
\end{equation}
Therefore, the system of equations (\ref{H-J-eq}) and (\ref{4}) is
now replaced by the system of equations (\ref{motion-eq-1}) and
(\ref{4}).

We look for a solution for $\rho$, assuming that $V$ is a linear
function of $z$, viz.
\begin{equation}\label{V-assumption}
V(z,s) = g(s)\,z + \kappa (s)\,
\end{equation}
and this corresponds to seek a quadratic form of the solution for
the phase $\theta (z,s)$, viz.
\begin{equation}\label{5}
\Theta \left(z,s\right) = \Theta_0\left(s\right) +
\kappa\left(s\right) z + \frac{1}{2}\,g(s)z^2,
\end{equation}
where the 'initial phase', $\Theta_0(s)$, the 'wavenumber',
$\kappa(s)$, and the 'dispersive' term, $g(s)$, are real functions
of the time-like variable $s$. It is easy to see that the arbitrary
function of $s$, $c_0(s)$, appearing in Eq. (\ref{motion-eq}), is
proportional to $\Theta_0'$ (hereafter the prime stands for the first-order derivative with respect to $s$), i.e. $c_0(s) = -\Theta_0'(s)$.

After substituting Eq. (\ref{V-assumption}) into Eq.
(\ref{motion-eq-1}), the system of equations
 (\ref{4}) and (\ref{motion-eq-1}) can be cast into the form
 \begin{equation}\label{continuity-1}
    \frac{\partial \rho}{\partial s} + \left(g\,z + \kappa \right)\frac{\partial\rho}{\partial z} + g\,\rho = 0\,,
\end{equation}
and
{\small\begin{eqnarray}\label{motion-eq-2}
&-&\rho\left(g'\,z + \kappa'\right) +\left(g\,z +
\kappa\right)\frac{\partial\rho}{\partial s} +
\left(-2\Theta_0' -g'\,z^2 - 2\kappa'\,z\right)\frac{\partial
\rho}{\partial z} \nonumber \\
&-& \omega_z^2\,z\rho
-\omega_z^2\,z^2\frac{\partial\rho}{\partial z} - 3 q_{1D}
\, \rho \frac{\partial \rho}{\partial z} +
\frac{\lambda_c^2}{4}\frac{\partial^3 \rho}{\partial z^3} = 0\,.
\end{eqnarray}}
Then, substituting Eq. (\ref{continuity-1}) into Eq.
(\ref{motion-eq-2}) we obtain
\begin{equation}\label{rho-eq}
    -\left(g'+g^2 + \omega_z^2\right)\left(\rho +z \frac{\partial\rho}{\partial z}\right)z -\left(\kappa' + g\,\kappa\right) \left(\rho + 2\,z\frac{\partial \rho}{\partial z}\right)
    - 2\theta_0'\frac{\partial \rho}{\partial z} - 3 q_{1D}\, \rho \frac{\partial \rho}{\partial z} + \frac{\lambda_c^2}{4}\frac{\partial^3 \rho}{\partial z^3} = 0\,,
\end{equation}
where $\theta_0'(s) \equiv \Theta_0'(s) + \kappa^2(s)/2$. To
reduce Eq. (\ref{rho-eq}) to the following KdV-like equation
\begin{equation}\label{70}
-2\theta_0'(s) \frac{\partial \rho}{\partial z} - 3 q_{1D}(s)\,
\rho \frac{\partial \rho}{\partial z} +
\frac{\lambda_c^2}{4}\frac{\partial^3 \rho}{\partial z^3} = 0\,,
\end{equation}
we have to impose that the coefficients of $\left(\rho +z
\frac{\partial\rho}{\partial z}\right)z$ and $\left(\rho +
2\,z\frac{\partial \rho}{\partial z}\right)$ are zero, namely we
automatically find that $g(s)$ satisfies the following Riccati's
equation, viz.
\begin{equation}\label{6}
g'+g^2 + \omega_z^2=0\,,
\end{equation}
while $\kappa(s)$ is related with it through
\begin{equation}\label{6a}
\kappa' + g\,\kappa =0\,,
\end{equation}which is readily integrated as
\begin{equation}\label{6b}
\kappa\left(s\right) = \kappa_0\,e^{-\int_0^s{g\left(\tau\right)
\,d\tau}},
\end{equation}
where $\kappa_0$ is an arbitrary constant.

We look for functions $\rho (z,s)$ which satisfies simultaneously
the KdV-like equation (\ref{70}) and the continuity equation
(\ref{continuity-1}). By using Eqs. (\ref{2}),(\ref{5}) and
(\ref{6})-(\ref{6b}), they allow us to construct also solutions of
the longitudinal equation (\ref{104}).

To this end, under the coordinate transformation
\begin{eqnarray}\label{transf}
    \xi &=& \xi(z,s) = q_{1D}(s)\,z + R(s)\nonumber\\
    &&\\
    s' &=& s'(z,s) = s\,,\nonumber
\end{eqnarray}
where $R(s)$ is a real function, the system of Eqs.
(\ref{continuity-1}) and (\ref{70}) becomes
\begin{equation}\label{continuity-2}
    \left[\left(\frac{q_{1D}'}{q_{1D}} + g\right)\left(\xi - R\right) + R' + \kappa\,q_{1D} \right]\frac{\partial\rho}{\partial\xi} + \frac{\partial \rho}{\partial s'} + g\,\rho = 0\,,
\end{equation}
and
\begin{equation}\label{rho-eq-1}
-2\theta_0' \frac{\partial \rho}{\partial \xi} - 3 q_{1D}\, \rho
\frac{\partial \rho}{\partial \xi} +
\frac{\lambda_c^2}{4}\,q_{1D}^2\,\frac{\partial^3 \rho}{\partial
\xi^3} = 0\,,
\end{equation}
where the prime denotes differentiation with
respect to $s$.

To find solutions in the factorized form
\begin{equation}\label{factorized-form}
\rho(\xi,s') = A(s')F(\xi)\,,
\end{equation}
satisfying simultaneously (\ref{continuity-2}) and
(\ref{rho-eq-1}), the following conditions have to be imposed
\begin{eqnarray}\label{conditions}
    q_{1D}' + g\, q_{1D} &=& 0\nonumber\\
    &&\\
    R' + \kappa\,q_{1D} &=&0\,.\nonumber
\end{eqnarray}
Consequently, Eqs. (\ref{continuity-2}) and (\ref{rho-eq-1}) become
the following ordinary differential equations, respectively,
\begin{equation}\label{continuity-3}
    A' + g\,A = 0\,,
\end{equation}
and
\begin{equation}\label{F-eq}
-2\theta_0' \frac{dF}{d\xi} - 3 q_{1D}\, A\,F \frac{dF}{d\xi} +
\frac{\lambda_c^2}{4}\,q_{1D}^2\,\frac{d^3 F}{d\xi^3} = 0\,.
\end{equation}
We note that the first condition (\ref{conditions}) and continuity
equation (\ref{continuity-3}) imply, respectively
\begin{equation}\label{G-solution}
    q_{1D}(s') = q_0\,e^{-\int_{0}^{s'}g\left(\tau\right)\,d\tau}\,,
\end{equation}
and
\begin{equation}\label{A-solution}
    A(s') = A_0\,e^{-\int_{0}^{s'}g\left(\tau\right)\,d\tau}\,,
\end{equation}
where $q_0$ and $A_0$ are integration constants, i.e. $q_0 =
q_{1D}(s'=0)$, $A_0 = A(s'=0)$. Additionally, by using solutions
(\ref{G-solution}) and (\ref{6b}), the second condition
(\ref{conditions}) can be easily solved for $R(s')$, viz.
\begin{equation}\label{R-solution}
    R(s') = R_0 - \kappa_0\,q_0\,\int_{0}^{s'}ds''e^{-2\int_{0}^{s''}g\left(\tau\right)\,d\tau}\,,
\end{equation}
where $R_0 = R(s'=0)$.

Furthermore, we also observe that, given the set of Eqs.
(\ref{6}), (\ref{6b}), (\ref{G-solution}) - (\ref{R-solution}), we
can conveniently express the functions $A(s)$, $k(s)$ and
$\omega_z (s)$ in terms of the controlling parameter $q_{1D}(s)$
(which is directly connected with the transverse part of the GPE
solution), i.e.
\begin{equation}\label{parameter-A}
A(s) = \frac{A_0}{q_0}\,q_{1D}(s)\,,
\end{equation}
\begin{equation}\label{parameter-kappa}
\kappa (s) = \frac{\kappa_0}{q_0}\,q_{1D}(s)\,,
\end{equation}
\begin{equation}\label{parameter-R}
R(s) = -\frac{\kappa_0}{q_0}\int_{0}^{s}\,q_{1D}^2 (\tau)\,d\tau +
R_0\,,
\end{equation}
\begin{equation}\label{parameter-omega-z}
\omega_z^2(s) = - q_{1D}
(s)\frac{d^2}{ds^2}\left[\frac{1}{q_{1D}(s)}\right]\,.
\end{equation}
In particular, Eq. (\ref{parameter-omega-z}) has been obtained by
substituting the first of equations (\ref{conditions}) into Riccati's
equation (\ref{6}). It establishes a 'control condition' by the
transverse part of the GPE solution on the longitudinal parabolic
potential. In fact, it indicates which time dependence of $\omega_z
= \omega_z (s)$ has to be taken, provided that $q_{1D}(s)$, given by
Eq. (\ref{q1D}), is solution of the first of Eqs. (\ref{conditions}).
Note that, according to Eq. (\ref{q1D}), it results that
\begin{equation}\label{q-0}
    q_0 = q\,\frac{
{\cal F}\left[\sigma_x (s), \sigma_y
(s)\right]_{s=0}}{\sigma_x\left(s=0\right)\sigma_y\left(s=0\right)}\,.
\end{equation}
(Note that ${\cal F}\left[\sigma_x (s), \sigma_y (s)\right]_{s=0}$
does not coincides with ${\cal F}\left[\sigma_x (s=0), \sigma_y
(s=0)\right]$.) To cast Eq. (\ref{F-eq}) as an equation with
constant coefficients, we can choose the arbitrary function
$\Theta'_0 (s')$ proportional to
$e^{-2\int_0^{s'}g\left(\tau\right) \,d\tau}$ and therefore
\begin{equation}\label{Theta-prime-zero}
    \Theta_0' (s') = \,\frac{\widetilde{\Theta}_0'}{q_0^2}q_{1D}^2(s')\,
\end{equation}
where $\widetilde{\Theta}_0'$ is an arbitrary constant, in such a
way that, according to Eq. (\ref{6b}) and the definition of
$\theta_0' (s')$ given above, we have
\begin{equation}\label{theta-prime-zero}
    \theta_0' (s') = \frac{\widetilde{\Theta}_0' + \kappa_0^2/2}{q_0^2}\,q_{1D}^2(s')
    \equiv \,\frac{\widetilde{\theta}_0'}{q_0^2}\,q_{1D}^2(s')\,.
\end{equation}
By substituting Eqs. (\ref{parameter-A}), (\ref{parameter-kappa})
and (\ref{theta-prime-zero}) in Eq. (\ref{F-eq}), we finally obtain the
following ordinary differential equations with constant coefficients
(stationary KdV equation)
\begin{equation}\label{stationary-KdV}
-2\widetilde{\theta}_0' \frac{dF}{d\xi} - 3 q_0\, A_0\,F
\frac{dF}{d\xi} + \frac{\lambda_c^2\,q_0^2}{4}\,\frac{d^3
F}{d\xi^3} = 0\,.
\end{equation}

Let us now determine the phase $\Theta(z,s)$. To this end, we
first observe that $\Theta_0 (s)$ can be obtained by integrating
Eq. (\ref{Theta-prime-zero}), i.e.
\begin{equation}\label{Theta-zero}
    \Theta_0 (s) = \varphi_0 + \frac{\widetilde{\Theta}_0'}{q_0^2}\,\int_{0}^{s}q_{1D}^2(\tau)\,d\tau\,,
\end{equation}
where $\varphi_0$ is an integration constant. Without loss of
generality, we can put: $\varphi_0 = 0$. Then, by using the first of
conditions (\ref{conditions}) and Eqs. (\ref{parameter-kappa}) and
(\ref{Theta-zero}), Eq. (\ref{5}) can be easily cast into the form
\begin{equation}\label{Theta-bis}
    \Theta (z,s) = -\frac{1}{2}\frac{q_{1D}'(s)}{q_{1D}(s)}\left[z-\zeta (s)\frac{}{}\right]^2 + \Delta (s)\,,
\end{equation}
where
\begin{equation}\label{greek-zeta}
    \zeta (s) = \frac{\kappa_0}{q_0}\frac{q_{1D}^2(s)}{q_{1D}'(s)}\,,
\end{equation}
\begin{equation}\label{Delta}
    \Delta (s) = \frac{\widetilde{\Theta}_0'}{q_0^2} \,\int_{0}^{s}q_{1D}^2 (\tau)\,d\tau
    + \frac{\kappa_0^2}{q_0^2}\frac{q_{1D}^3(s)}{q_{1D}'(s)}\,.
\end{equation}

\subsubsection{Soliton solutions \label{soliton solutions}}

As it is well known, Eq. (\ref{stationary-KdV}) admits both
localized and periodic solutions \cite{Sulem99,Whitham99,Dau06}. Typically, the latter
are expressed in terms of Jacobian elliptic functions, whose
suitable asymptotic limits of their parameters recover the
localized solutions in the form of bright, dark and grey solitons.
However, a very useful integration approach of Eq.
(\ref{stationary-KdV}) that has a simple physical meaning of the
solutions is the well-known Sagdeev's pseudo-potential method
\cite{Kadomtsev,Karpman}.

\medskip\medskip

\noindent (i). \underline{Bright solitons}\\

\medskip

If $F$ (and consequently $\psi_z$) is a normalizable wave
function, we can look for a bright soliton solution of Eq.
(\ref{stationary-KdV}), which satisfies the following boundary
conditions: $F\,\rightarrow\, 0$, for $\xi\,\rightarrow\,\pm
\infty$. This solution exists for $\widetilde{\theta}_0'
>0$\,, \cite{FedeleSchamel02}, i.e.
\begin{equation}\label{F-bright}
F(\xi) = -\frac{2\widetilde{\theta}_0'}{q_0 A_0}\,\mbox{sech}^2
\left(\frac{\sqrt{2\widetilde{\theta}_0'}}{\lambda_c
|q_0|}\,\xi\right)\,.
\end{equation}
Then, going back to the old variables, $z$ and $s$ and using Eqs.
(\ref{factorized-form}),  (\ref{parameter-A}) and
(\ref{parameter-R}), we finally get the following bright soliton
solution of Eq. (\ref{70})
\begin{equation}\label{rho-bright}
    \rho(z,s) = -\frac{2\widetilde{\theta}_0'}{q_0^2}\,q_{1D}(s)\,\,\mbox{sech}^2
\left\{\frac{\sqrt{2\widetilde{\theta}_0'}}{\lambda_c
|q_0|}q_{1D}(s)\left[z - z_0(s)\right]\right\}\,,
\end{equation}
where
\begin{equation}\label{z-zero}
z_0 (s) = \frac{1}{q_{1D}(s)}\left[\frac{\kappa_0
}{q_0}\int_{0}^{s}q_{1D}^2(\tau)\,d\tau - R_0\right]\,.
\end{equation}
The positivity of $\rho (z,s)$ implies that $q_{1D} <0$ and
therefore $q <0$ and $q_0 <0$. On the other
hand, if we require that $\psi_z$ is normalized, i.e.
\begin{equation}\label{parnorm}
\int dz\, |\psi_z|^2 = 1\,,
\end{equation}
then
\begin{equation}\label{z-normalization condition}
    \widetilde{\theta}_0' = \frac{q_0^2}{8} = \frac{q^2}{8}\,\left[\frac{
{\cal F}\left[\sigma_x (s), \sigma_y (s)\right]_{s=0}}
{\sigma_x\left(s=0\right)\sigma_y\left(s=0\right)}\right]^2\,.
\end{equation}

It follows that, in the case of bright solitons, Eq. (\ref{Delta})
becomes
\begin{equation}\label{Delta-bis}
    \Delta (s) = \frac{1}{2q_0^2}\left(\frac{q_0^2}{4} - \kappa_0^2\right)\,\int_{0}^{s}q_{1D}^2 (\tau)\,d\tau
    + \frac{\kappa_0^2}{q_0^2}\frac{q_{1D}^3(s)}{q_{1D}'(s)}\,.
\end{equation}

\medskip\medskip

\noindent(ii). \underline{Grey and dark solitons}\\

\medskip

If $F$ (and consequently $\psi_z$) is a non-normalizable solution,
we can look for dark or grey soliton solutions of Eq.
(\ref{stationary-KdV}), which satisfy the following boundary
conditions: $F\,\rightarrow\, F_0$, for $\xi\,\rightarrow\,\pm
\infty$, where $|F_0|\,< \,\infty$. These solutions are given by the
general form
\begin{equation}\label{grey-dark}
F (\xi) =
F_0\,\left[1-\epsilon^2\,\mbox{sech}^2\left(\xi/\Delta_0\right)\right]\,,
\end{equation}
where $\epsilon$ is a real parameter,
\begin{equation}\label{F0}
    F_0 = \frac{2\widetilde{\theta}_0'}{q_0A_0\left(\epsilon^2 -3\right)}\,,
\end{equation}
and
\begin{equation}\label{Delta0}
    \Delta_0 = \frac{\lambda_c^2 q_0^2}{2\widetilde{\theta}_0'}\left(1 - \frac{3}{\epsilon^2}\right)\,.
\end{equation}
We first observe that since $\Delta_0 >0$, the following condition holds:
\begin{equation}\label{inequality}
    \widetilde{\theta}_0' \left(\epsilon^2 -3\right) >0.
\end{equation}
Taking into account Eqs. (\ref{factorized-form}) and
(\ref{grey-dark})-(\ref{Delta0}), we easily get:
\begin{equation}\label{rho-grey-dark}
   \rho\left(\xi,s'\right) = \frac{2\widetilde{\theta}_0'\,q_{1D}(s')}{q_0^2\left(\epsilon^2 -3\right)}\left[1 - \epsilon^2\,\mbox{sech}^2\left(\sqrt{\frac{2\widetilde{\theta}_0'}{1- 3/\epsilon^2}}\,\,\frac{\xi}{\lambda_c |q_0|}\right)\right]\,,
\end{equation}
which, going back to the variables $z$ and $s$, can be cast into the form
\begin{equation}\label{rho-grey-dark-1}
    \rho\left(z,s\right) = \frac{2\widetilde{\theta}_0'\,q_{1D}(s)}{q_0^2\left(\epsilon^2 -3\right)}\left\{1 - \epsilon^2\,\mbox{sech}^2\left[\sqrt{\frac{2\widetilde{\theta}_0'}{1- 3/\epsilon^2}}\,\,\frac{q_{1D}(s)}{\lambda_c |q_0|}\left(z - z_0(s)\,\right)\right]\right\}\,.
\end{equation}
Taking into account condition (\ref{inequality}), from non-negativity of 
$\rho (z,s)$ it follows that $q_{1D}>0$. By virtue of Eq. (\ref{q1D}), this 
condition implies, in turn, that $q > 0$ and, consequently, due to Eq. (\ref{q-0}), that $q_0 >0$.

\begin{itemize}
\item If we choose $\epsilon = \pm 1$, then Eq. (\ref{inequality}) implies 
that $\widetilde{\theta}_0' <0$, while Eqs. (\ref{grey-dark}) and (\ref{rho-grey-dark-1}) take the form of standard 'dark solitons'. i.e.
\begin{equation}\label{dark}
F (\xi) = -\frac{\widetilde{\theta}_0 '}{A_0}\,\mbox{tanh}^2
\left[\frac{\sqrt{-\widetilde{\theta}_0 '}}{\lambda_c |q_0|}\,\xi
\right]\,,
\end{equation}
and
\begin{equation}\label{rho-dark}
    \rho\left(z,s\right) = -\frac{\widetilde{\theta}_0'\,q_{1D}(s)}{q_0^2} \,\mbox{tanh}^2\left[\sqrt{-\widetilde{\theta}_0'}\,\,
\frac{q_{1D}(s)}{\lambda_c |q_0|}\left(z - z_0(s)\,\right)\right]\,.
\end{equation}
\item For $\epsilon \neq \pm 1$, condition (\ref{inequality}) and
Eq. (\ref{rho-grey-dark-1}) take the form of standard
'grey solitons', in the following range of the parameters
$\epsilon$ and $\widetilde{\theta}_0'$, i.e.
\begin{equation}\label{grey-ranges}
-1< \epsilon < 1~~~\mbox{and}\,\,\,\,\,\widetilde{\theta}_0' <
0\,.
\end{equation}

\end{itemize}

\section{The evolution and collapse of the controlled GPE structures}\label{collapse}

In this section, we carry out a stability analysis of our system,
taking into account the control condition (\ref{parameter-omega-z})
and the explicit dependence of the controlling parameter $q_{1D}$ on
the transverse scale lengths $\sigma_x$ and $\sigma_y$ through Eq.
(\ref{q1D}). For simplicity, we present the results for the simple
case when the perpendicular solution is the product of only two
single Hermite-Gauss modes, which implies for $q_{1D}$ the
expression (\ref{q1D-bis}) for the transverse mode $(n,l)$. The
generalization to the multiple modes' solutions may be somewhat
lengthy, but is straightforward.

We note that the characteristic spatial scales $\sigma_x$,
$\sigma_y$, and $q_{1D}^{n,l}(s)$ are related with the coefficients
of the restoring force $\omega_x$, $\omega_y$, and $\omega_z$,
through four equations. These are the Ermakov-Pinney equations
(\ref{Pinnerm}), the minimization condition for the control, Eq.
(\ref{q1D-bis}), and the consistency condition of the amplitude and
the phase of the parallel solution, Eq. (\ref{parameter-omega-z}),
\begin{equation}\label{Pinnerm-2}
\frac{d^2\sigma_j\left(s\right)}{ds^2} + \omega_j^2(s)
 \sigma_j\left(s\right) -\frac{\lambda_c^2
}{4\,\sigma_j\left(s\right)^3 } = 0,\quad j=x,y,
\end{equation}
\begin{equation}\label{q1D
-terc} q_{1D}^{n,l}\left(s\right) = q\,\frac{
\delta_n\,\delta_l}{\sigma_x\left(s\right)\sigma_y\left(s\right)},
\end{equation}
\begin{equation}\label{22-bis}
\omega_z^2 \left(s\right) = -q_{1D}^{n,l}\frac{d^2}{d s^2}
\left(\frac{1}{q_{1D}^{n,l}}\right).
\end{equation}
Obviously, if we wish to produce a controlled state in a BEC
experiment, only two coefficients of the restoring force can be
adopted arbitrarily, while the third is determined by the above
constraints. Alternatively, we may adopt two arbitrary dependencies
in the form $F_1(\omega_x, \omega_y, \omega_z, s) = 0$,
$F_2(\omega_x, \omega_y, \omega_z, t) = 0$, from which the
quantities $\omega_j(s),\, j = x,y,z$ are uniquely determined with
the use of the constraints (\ref{Pinnerm-2})-(\ref{22-bis}). The
temporal evolution and the stability of the resulting coherent
nonlinear state depends strongly on our choice of the temporal
dependence of the confining potential $V_{trap}$. Both stable and
collapsing solutions can be obtained under different conditions, which
is studied in more details below.

\subsection{Stable solutions}

First we study the important particular case when the {\it
perpendicular} restoring force is stationary. 
For this case, we demonstrate the existence of stable (non-collapsing) nonlinear
modes when the perpendicular solution contains only one
Gauss-Hermite mode in each direction, see Eq. (\ref{superposition}).

The Ermakov-Pinney equation (\ref{Pinnerm}) is easily solved if
$\omega_j = {\rm constant}$, where $j=x,y$. The general solution for
the perpendicular scale and for the phase functions, Eqs.
(\ref{gamma}) and (\ref{phi}), can be written as
\begin{equation}\label{sigmadva}
\sigma_j\left(s\right) =  \frac{\sqrt{4 \gamma_{j,0}^2
\sigma_{j,0}^4+4 \omega_j^2
   \sigma_{j,0}^4+8 \gamma_{j,0} \omega_j \sigma_{j,0}^4 \sin (2
   \omega_j s) +\lambda_c^2-\left[4
   \left(\gamma_{j,0}^2-\omega_j^2\right) \sigma_{j,0}^4 +\lambda_c^2\right] \cos (2 \omega_j s )}}{2
   \sqrt{2}\, \sigma_{j,0} \omega_j},
\end{equation}
\begin{equation}\label{gammadva}
 \gamma_j\left(s\right) =  \frac{\omega_j \left\{8 \gamma_{j,0} \omega_j
  \sigma_{j,0}^4 \cos (2\omega_j s )+\left[4 \left(\gamma_{j,0}^2-\omega_j^2\right) \sigma_{j,0}^4+\lambda_c^2\right] \sin (2
\omega_j s )\right\}}{4 \gamma_{j,0}^2
   \sigma_{j,0}^4+4 \omega_j^2 \sigma_{j,0}^4+8
   \gamma_{j,0} \omega_j \sigma_{j,0}^4 \sin (2 \omega_j s )
   +\lambda_c^2-\left[4 \left(\gamma_{j,0}^2 -\omega_j^2\right) \sigma_{j,0}^4+\lambda_c^2 \right] \cos (2 \omega_j s )},
\end{equation}
and
\begin{equation}\label{phidva}
 \phi_{j,k}\left(s\right) =  \phi_{j,k,0}-\frac{2 k+1}{2} \left\{\tan^{-1}\left[\frac{2 \gamma_{j,0}
\sigma_{j,0}^2}{\lambda_c}+\frac{4 \gamma_{j,0}^2 \sigma_{j,0}^4
+\lambda_c^2}{2 \lambda_c \sigma_{j,0}^2 \omega_j }\,\tan \left(
   \omega_j s\right)\right] -
   \tan^{-1}\left(\frac{2 \gamma_{j,0}
\sigma_{j,0}^2}{\lambda_c}\right)
   \right\},
\end{equation}
where $\sigma_{j,0} = \sigma_j(0)$, $\gamma_{j,0} = \gamma_j(0)$,
and , $\phi_{j,k,0} = \phi_{j,k}(0)$ are arbitrary initial
conditions.

We now restrict ourselves to a 2D geometry, $\partial/\partial y = 0$,
where the analysis is particularly simple, but it can be easily
generalized also to the 3D case. In a 2D case we have
\[
z_0\left(s\right) =
\frac{1}{q_{1D}\left(s\right)}\left(-R_0+\frac{2\,\delta_k^2}{q_0
\lambda_c}\left\{\tan^{-1}\left[\frac{2 \gamma_{x,0}
\sigma_{x,0}^2}{\lambda_c}+\frac{4 \gamma_{x,0}^2 \sigma_{x,0}^4
+\lambda_c^2}{2 \lambda_c \sigma_{x,0}^2 \omega_x }\,\tan \left(
\omega_x s\right)\right] - \right.\right.
\]
\begin{equation}\label{z0dva}
\left.\left. \tan^{-1}\left(\frac{2 \gamma_{x,0}
\sigma_{x,0}^2}{\lambda_c}\right)\right\}\right),
\end{equation}
and we also conveniently rewrite the expression for $\sigma_x$, Eq.
(\ref{sigmadva}), in the following form
\begin{equation}\label{sigmadrugo}
\sigma^2_x\left(s\right)  =  D_x^2\, \left\{2\,\cos^2\left[ \omega
_x \left(s-S_x\right)\right]-1 + \sqrt{1 + \frac{\lambda_c^2}{4
\omega_x^2 D_x^4}}\right\},
\end{equation}
where $D_x$ and $S_x$ are two arbitrary constants. It is obvious
that $\sigma_x^2(s)$ is non-negative, and that it may have periodic
zeros only in the absence of spatial dispersion, $\lambda_c = 0$.
Thus, the amplitude and the characteristic wavenumber of the
solution, $\propto q_{1D}\propto 1/ \sigma_x(s)$ are regular
functions, and the collapse does not occur. The corresponding
variation of the coefficient of the parallel restoring force,
$\omega_z$, in a 2D regime with a single Hermite-Gauss mode, using
Eq. (\ref{parameter-omega-z}), $\omega_z$ is readily obtained as

\begin{eqnarray}\label{omega_z_2-D}
\nonumber
   \frac{\omega_z^2\left(s\right)}{\omega_x^2}&=& 1 -
\frac{\lambda_c^2}{4\omega_x^2 D_x^4}\, \left\{2\,\cos^2\left[\omega
_x \left(s-S_x\right)\right]-1 + \sqrt{1 + \frac{\lambda_c^2}{4
\omega_x^2 D_x^4}}\right\}^{-2} \\
\nonumber
  &=&\frac{\cos\left[2 \omega _x \left(s-S_x\right)\right]+ \sqrt{1 +
{\lambda_c^2}/\left({4 \omega_x^2 D_x^4}\right)}+\lambda_c/\left({2
\omega_x D_x^2}\right)}{\left\{\cos\left[2 \omega _x
\left(s-S_x\right)\right] + \sqrt{1 + {\lambda_c^2}/\left({4
\omega_x^2 D_x^4}\right)}\right\}^2} \times\\
&& \left[\cos\left[2 \omega _x
\left(s-S_x\right)\right]+\frac{1}{\sqrt{1 + {\lambda_c^2}/\left({4
\omega_x^2 D_\bot^4}\right)}+\lambda_c/\left({2\omega_x
D_x^2}\right)}\right].
\end{eqnarray}
The first term is positively definite, and the second is the sum of a
harmonic function and a constant, of the form $\cos[2\omega_\bot
(s-S_\bot)] + \zeta$, where $0<\zeta<1$, which obviously features
the change of sign.
\begin{figure}[htb]
\centering
\includegraphics[width=160mm]{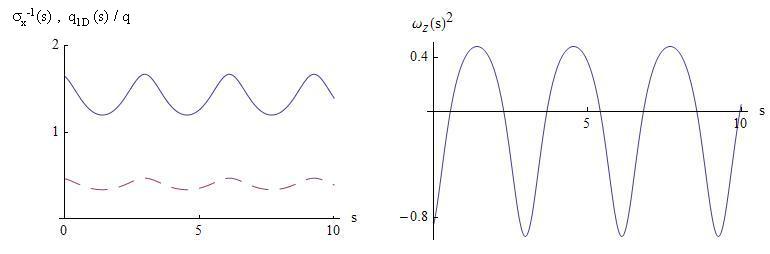}
\caption{{\bf Left:} The temporal evolution ("breathing") of the
perpendicular scale $\sigma_x^{-1}$ (solid line) and the parallel
scale $q_{1D}(s)/q$ (dashed), found in a stable configuration, with
$|q_0| = 1, \lambda_c = 1, \delta \equiv \lambda_c |q_0|/\sqrt{2
|\widetilde{\theta_0'}|} = 1, \omega_x = 1, \sigma_{x,0}=0.607107,
\gamma_{x,0} = .15, \phi_{x,k,0} = 0,$ and for the lowest order
Hermite-Gauss mode, $ k = 0$. {\bf Right:} The corresponding solution for the parallel characteristic frequency $\omega_z^2(s)$ in the 2D case, given by
Eq. (\ref{omega_z_2-D}). There is a periodic change of sign, from
focussing to defocusing and vice versa.}\label{fig20}
\end{figure}

The temporal evolution ("breathing") of the
perpendicular scale $\sigma_x^{-1}$ and the corresponding evolution
of $\omega_z^2(s)$ in the 2D case are shown in Fig. \ref{fig20}, respectively.
Hereafter, we fix for all figures: $|q_0|=1$.

\begin{figure}[htb]
\centering
\includegraphics[width=160mm]{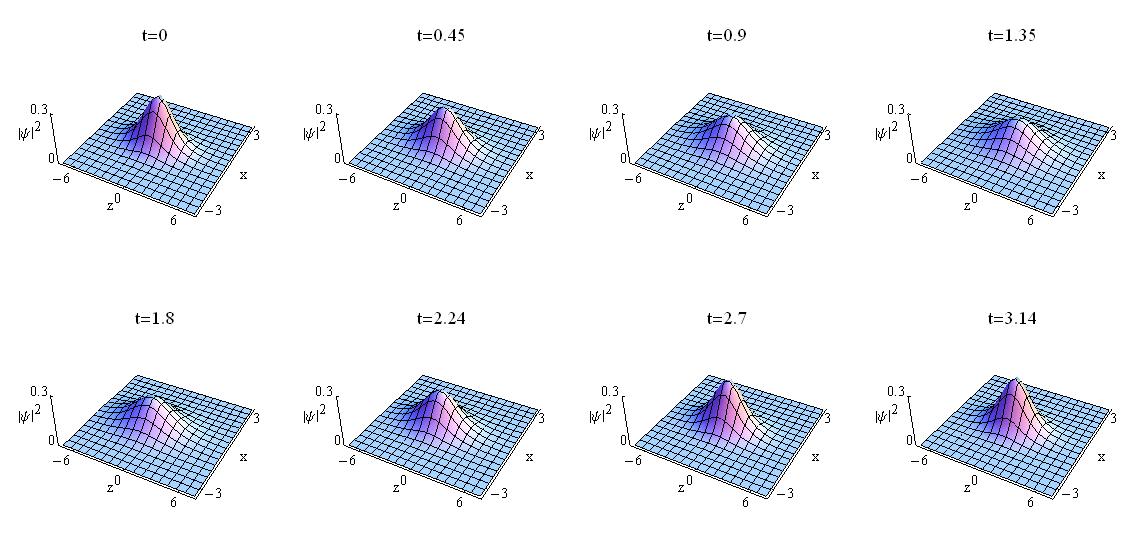}
\caption{The evolution of the lowest order ($k=0$), 2D controlled,
stable BEC state whose parallel component is a bright soliton. The
initial position is calculated with $R_0 = 0.2$, and the other
parameters of the solution are the same as in Fig. \ref{fig20}.
}\label{bright_k=0}
\end{figure}

\begin{figure}[htb]
\centering
\includegraphics[width=160mm]{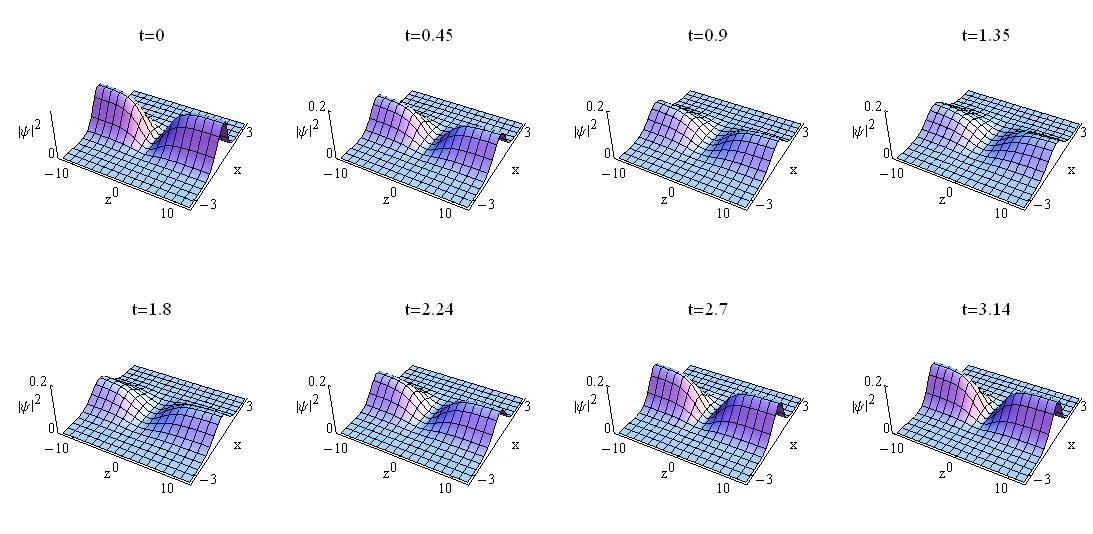}
\caption{The lowest order ($k=0$), 2D controlled, stable BEC state
whose parallel component is a dark soliton. The parameters of the
solution are the same as in Fig. \ref{fig20}}\label{dark_k=0}
\end{figure}

\begin{figure}[htb]
\centering
\includegraphics[width=160mm]{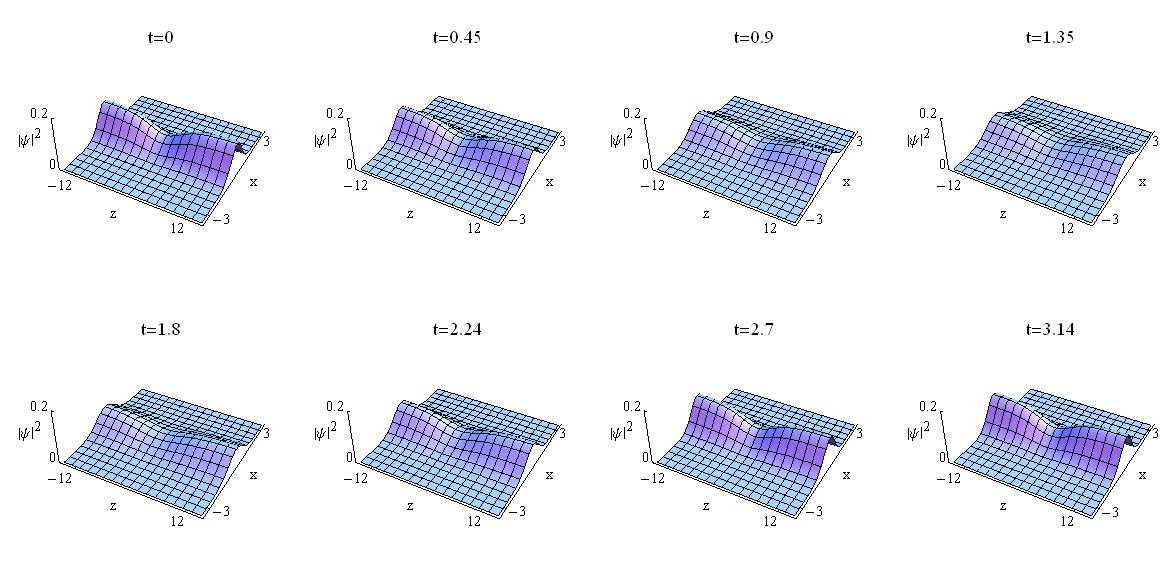}
\caption{The lowest order ($k=0$), 2D controlled, stable BEC state
that in the parallel direction behaves as a gray soliton with
$\epsilon = 0.7$. Other parameters are the same as in Fig.
\ref{fig20}}\label{gray_k=0}
\end{figure}

The evolution of the 2D structures that in the parallel direction
correspond to a bright, dark, and gray soliton, described by Eqs.
(\ref{rho-bright}), (\ref{dark}), and (\ref{rho-grey-dark-1}),
respectively, is for the case of a ground Hermite-Gauss mode, $k=0$
displayed in Figs. \ref{bright_k=0}-\ref{gray_k=0}. The solution
features "breathing" (the oscillations of the amplitude and of
the position) due to the oscillations of the perpendicular solution,
as well as oscillations in the parallel direction, described by
the quantity $z_0(s)$, Eq. (\ref{z-zero}). The frequency of the
latter oscillations is smaller than that of the "breathing", and
they are not noticeable within the time span of Figs.
\ref{bright_k=0}-\ref{gray_k=0}.

\begin{figure}[htb]
\centering
\includegraphics[width=160mm]{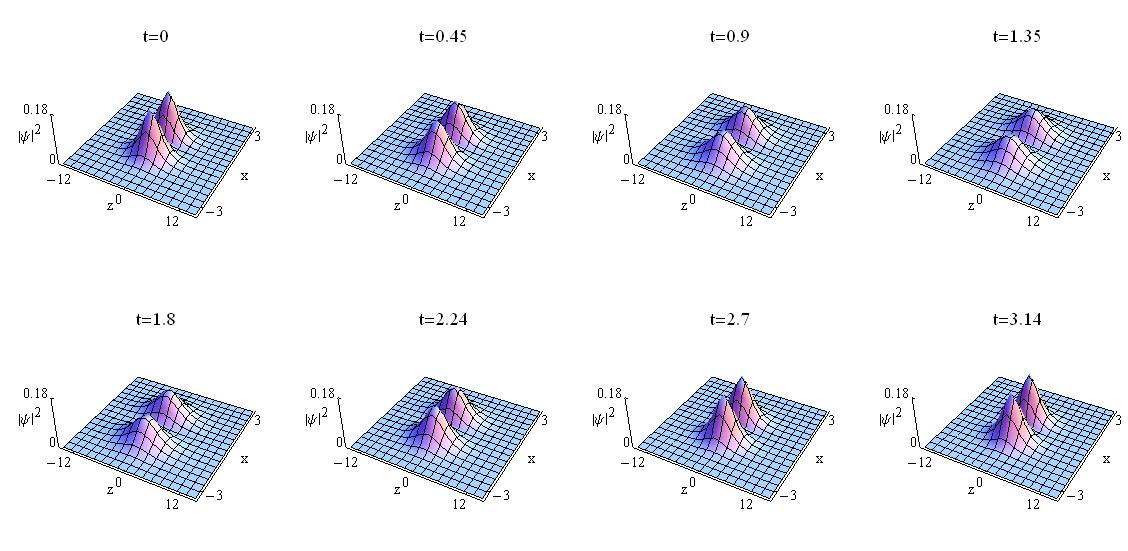}
\caption{The evolution of the first order ($k=1$), 2D controlled,
stable BEC state whose parallel component is a bright soliton. The
parameters of the solution are the same as in Figs.
\ref{bright_k=0}-\ref{gray_k=0}.}\label{bright_k=1}
\end{figure}

\begin{figure}[htb]
\centering
\includegraphics[width=160mm]{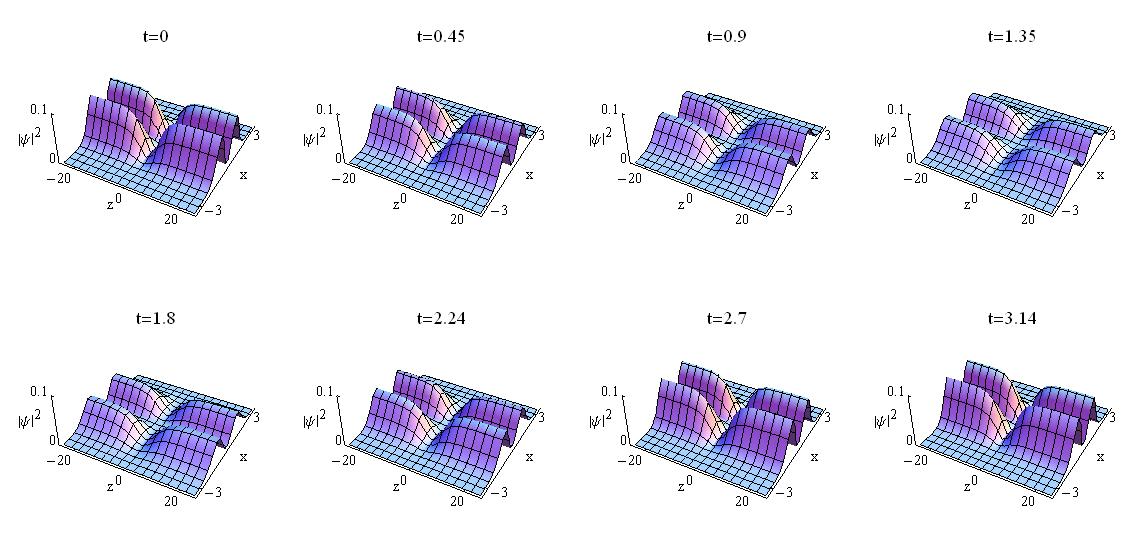}
\caption{The first order ($k=1$), 2D controlled, stable BEC state
whose parallel component is a dark soliton. The parameters of the
solution are the same as in Figs.
\ref{bright_k=0}-\ref{gray_k=0}.}\label{dark_k=1}
\end{figure}

\begin{figure}[htb]
\centering
\includegraphics[width=160mm]{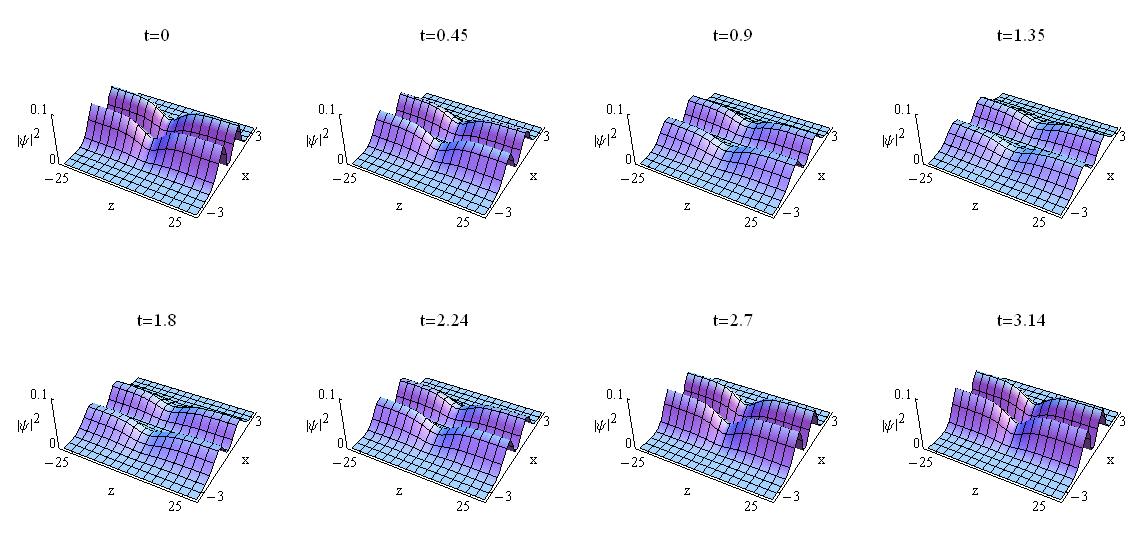}
\caption{The lowest order ($k=1$), 2D controlled, stable BEC state
that in the parallel direction behaves as a gray soliton with
$\epsilon = 0.7$. Other parameters are the same as in Figs.
\ref{bright_k=0}-\ref{gray_k=0}.}\label{gray_k=1}
\end{figure}

The evolution of the stable, first order (k=1) mode, is displayed in
Figs. \ref{bright_k=1}-\ref{gray_k=1}. Its behavior is similar to
that of the ground mode.

\subsection{Collapsing solutions}

Besides the stable solutions described in the preceding subsection,
unstable controlled BEC states can also be generated, with a
different choice of the confining potential $V_{trap}$. This can be
demonstrated in the simple case when the parallel restoring force is
adopted to be stationary, $\omega_z = {\rm constant}$, when the
equation (\ref{parameter-omega-z}) can be solved as
\begin{equation}\label{16a}
q_{1D}\left(s\right) = \frac{1}{c_1\cos\left( \omega_{ z} s +
\varphi_0 \right)},
\end{equation}
where $c_1$ and $\varphi_0$ are arbitrary constants. Then, in the
2D case ($\partial/\partial y =0$) and using Eqs. (\ref{q1D-bis}),
(\ref{Pinnerm}), (\ref{gamma}), (\ref{phi}), and (\ref{z-zero}), we
can readily write down the parameters of the perpendicular, Eq.
(\ref{HermiteGauss}), and parallel, Eqs. (\ref{rho-bright}),
(\ref{dark}), and (\ref{rho-grey-dark-1}), solutions, viz.
\begin{equation}\label{sigma-collapse}
\sigma_x\left( s\right) = \frac{\sigma_{x,0}}{\omega_{
z}}\left[\omega_{ z} \cos\left(\omega_{ z}s\right) +\gamma_{x,0}\sin
\left(\omega_{ z}s\right)\right],
\end{equation}
\begin{equation}\label{gamma-collapse}
\gamma_x\left(s\right)=\omega_{ z} \,\,\frac{\gamma_{x,0} \cos
\left( \omega_{ z} s\right)-\omega_{ z} \sin \left(\omega_{ z} s
\right)}{\gamma_{x,0} \sin \left(\omega_{ z} s\right)+\omega_{ z}
\cos \left(\omega_{ z}s\right)},
\end{equation}
\begin{equation}\label{phi-collapse}
\phi_{x,k}\left(s\right)= \phi_{x,k,0} - \frac{\lambda_c\left(1+2
k\right) }{4\,\sigma_{x,0}^2} \frac{\sin \left(\omega_{ z} s
\right)}{\omega_{ z} \cos \left(\omega_{ z}s \right) + \gamma_{x,0}
\sin \left(\omega_{ z} s\right)},
\end{equation}
\begin{equation}\label{omega_x-collapse}
\omega_x\left(s\right) = \omega_{ z}\,\sqrt{1 +
\frac{\lambda_c^2\omega_{ z}^2}{4\,\sigma_{x,0}^4}\, \left[\omega_{
z} \cos \left(\omega_{ z}s \right) + \gamma_{x,0} \sin
\left(\omega_{ z} s\right) \right]^{-4}},
\end{equation}
\begin{equation}\label{q1D-collapse}
q_{1D}\left(s\right) = \frac{\delta_k\omega_{
z}}{\sigma_{x,0}}\left[\omega_{ z} \cos \left(\omega_{ z}s \right) +
\gamma_{x,0} \sin \left(\omega_{ z} s\right)\right]^{-1},
\end{equation}
\begin{equation}\label{z_0-collapse1}
z_0\left(s\right) = -\frac{R_0}{q_{1D}\left(s\right)} +
\frac{\delta_k \kappa_0}{q_0 \sigma_{x,0} \omega_{ z}}\,
\sin\left(\omega_{ z} s\right),
\end{equation}
where $\delta_k$ is defined in Eq. (\ref{functional-F-bis}), while
$\sigma_{x,0}=\sigma_x(0)$, $\gamma_{x,0}=\gamma_x(0)$,
$\phi_{x,k,0}=\sigma_{x,k}(0)$, and $R_0$ are arbitrary initial
conditions.

We note that the characteristic wavenumbers and the amplitudes of
both the perpendicular and the parallel components of the BEC
controlled structure are periodic functions of time, with the
frequency $\omega_{ z}$, which take infinite values twice during
each period. The displacement from the equilibrium position,
$z_0(s)/q_{1D}(s)$, oscillates with the same frequency, but remains
limited. Such behavior corresponds to a cyclic collapse and recovery
of the soliton-like structure in both directions. However, the
recovery does not occur in reality. Our basic GPE is no
longer valid for large amplitudes, such as those associated with a
collapsing solution, and the new physical effects that arise under
such circumstances are likely to perturb the system in a way that
prevents its recurrence.

This kind of collapse is well known for the bright solitons, Eq.
(\ref{rho-bright}), in the case of a focussing nonlinear term in the
GPE, $q_{1D}<0$ and $\widetilde{\theta_0'}>0$. Our analysis
shows that the collapse occurs also for a defocusing nonlinearity,
$q_{1D}>0$ and $\widetilde{\theta_0'} < 0$, when the nonlinear
solution is periodic in $z$. The wavelength of such train of
nonlinear structures is proportional to $1/\sqrt{q_{1D}}$, and
becomes infinitesimally small in the singularities of $q_{1D}$, i.e.
the train of coherent structures undergoes a collapse.

\begin{figure}[htb]
\centering
\includegraphics[width=160mm]{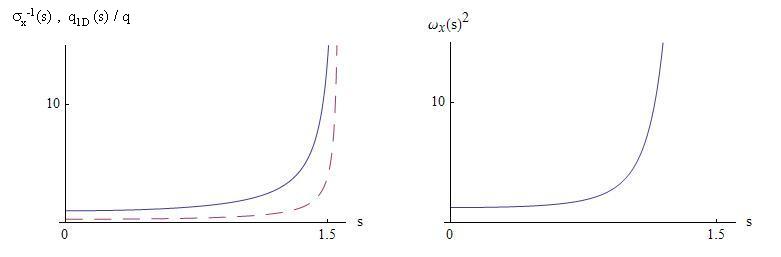}
\caption{{\bf Left:} The temporal evolution of the perpendicular
scale $\sigma_x^{-1}$ (solid line) and the parallel scale
$q_{1D}(s)/q$ (dashed), found in a collapsing configuration, with $|q_0|
= 1, \lambda_c = 1, \delta \equiv \lambda_c |q_0|/\sqrt{2
|\widetilde{\theta_0'}|} = 1, \omega_x = 1, \sigma_{x,0}=0.607107,
\gamma_{x,0} = .15, \phi_{x,k,0} = 0,$ and for the lowest order
Hermite-Gauss mode, $ k = 0$. {\bf Right:} The corresponding solution for the perpendicular
characteristic frequency $\omega_x^2(s)$ in the 2D case. Note that
the collapse needs to be supported by a rapid (explosive) growth of
the restoring force.}\label{fig30}
\end{figure}

\begin{figure}[htb]
\centering
\includegraphics[width=160mm]{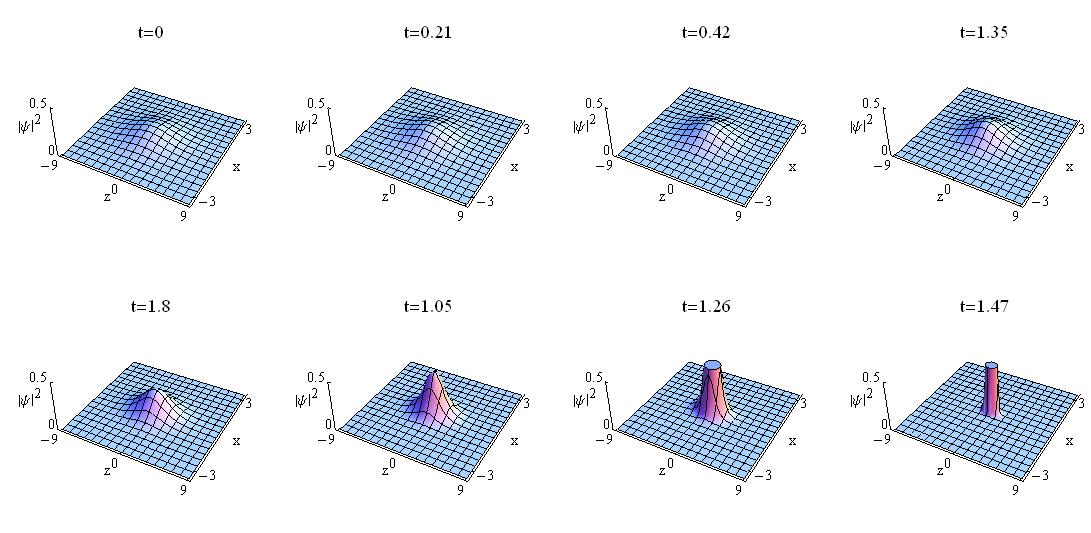}
\caption{The evolution of the lowest order ($k=0$), 2D controlled,
unstable BEC state whose parallel component is a bright soliton. The
initial position is calculated with $R_0 = 0.2$, and the other
parameters of the solution are the same as in Fig. \ref{fig30}.
}\label{unstable_bright_k=0}
\end{figure}

\begin{figure}[htb]
\centering
\includegraphics[width=160mm]{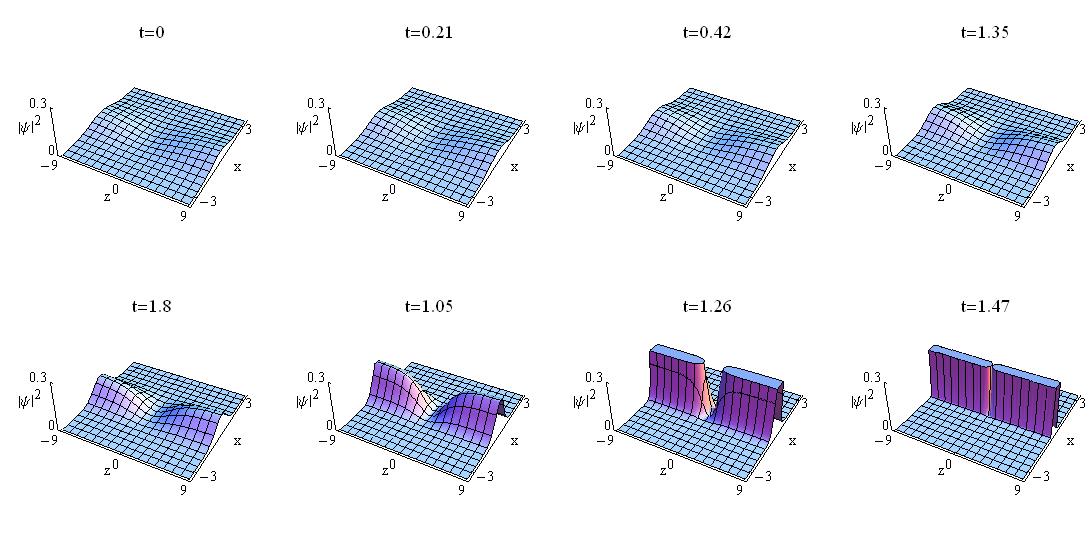}
\caption{The lowest order ($k=0$), 2D controlled, unstable BEC
state whose parallel component is a dark soliton. The parameters of
the solution are the same as in Fig.
\ref{fig30}}\label{unstable_dark_k=0}
\end{figure}

\begin{figure}[htb]
\centering
\includegraphics[width=160mm]{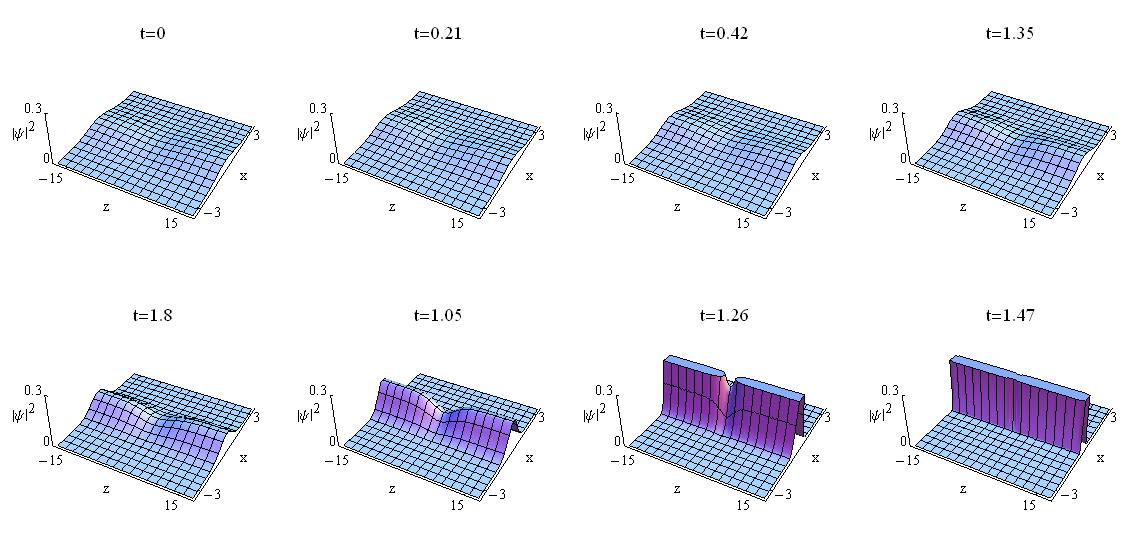}
\caption{The lowest order ($k=0$), 2D controlled, unstable BEC
state that in the parallel direction behaves as a gray soliton with
$\epsilon = 0.7$. Other parameters are the same as in Fig.
\ref{fig30}}\label{unstable_gray_k=0}
\end{figure}

\begin{figure}[htb]
\centering
\includegraphics[width=160mm]{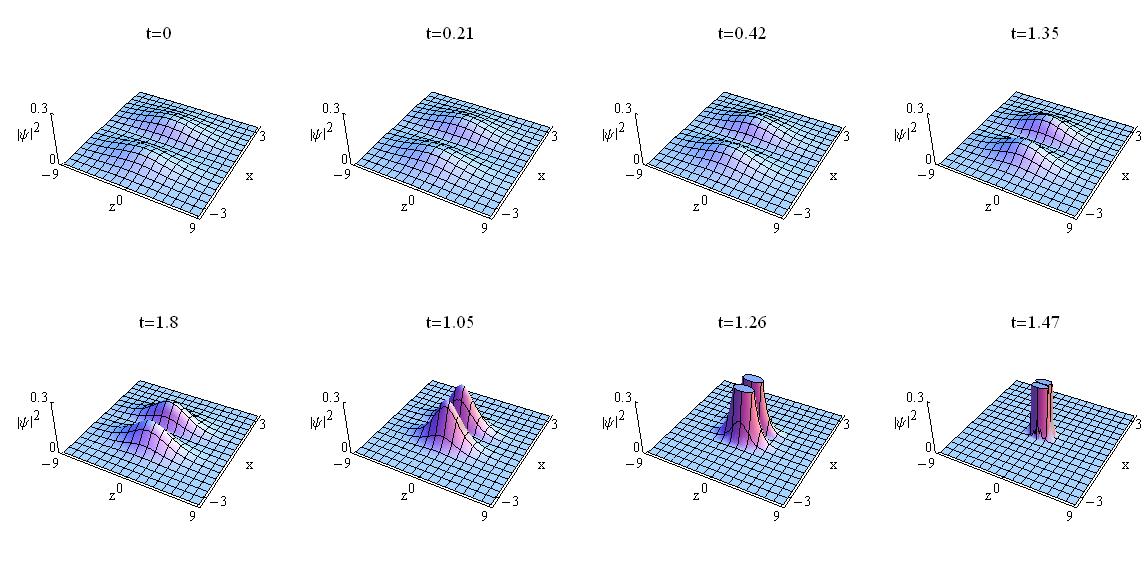}
\caption{The evolution of the first order ($k=1$), 2D controlled,
unstable BEC state whose parallel component is a bright soliton. The
parameters of the solution are the same as in Figs.
\ref{unstable_bright_k=0}-\ref{unstable_gray_k=0}.}\label{unstable_bright_k=1}
\end{figure}

\begin{figure}[htb]
\centering
\includegraphics[width=160mm]{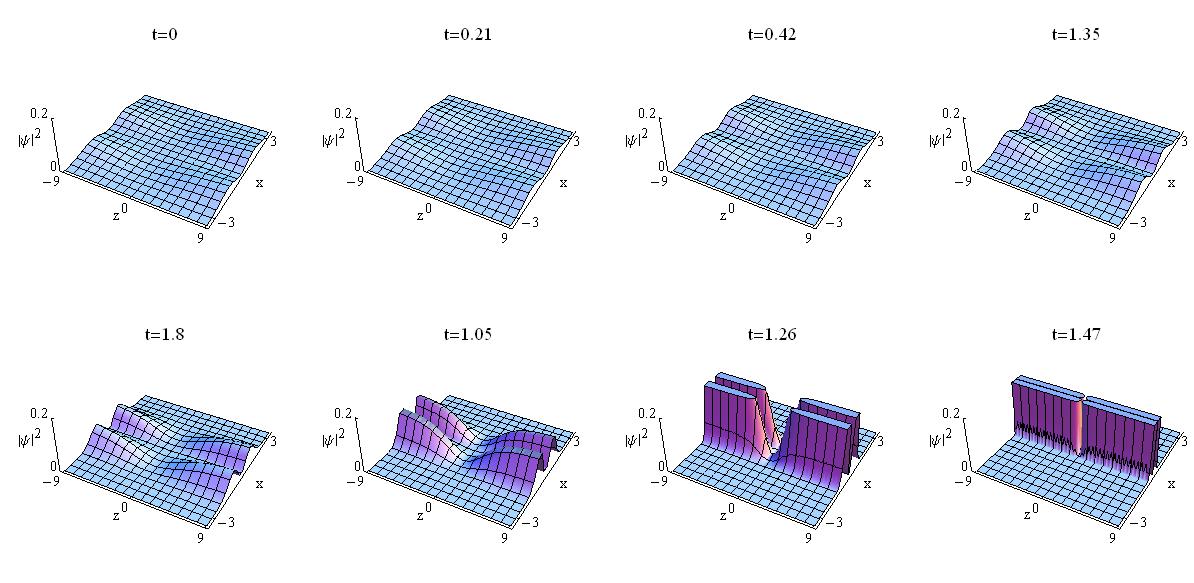}
\caption{The first order ($k=1$), 2D controlled, unstable BEC state
whose parallel component is a dark soliton. The parameters of the
solution are the same as in Figs.
\ref{unstable_bright_k=0}-\ref{unstable_gray_k=0}.}\label{unstable_dark_k=1}
\end{figure}

\begin{figure}[htb]
\centering
\includegraphics[width=160mm]{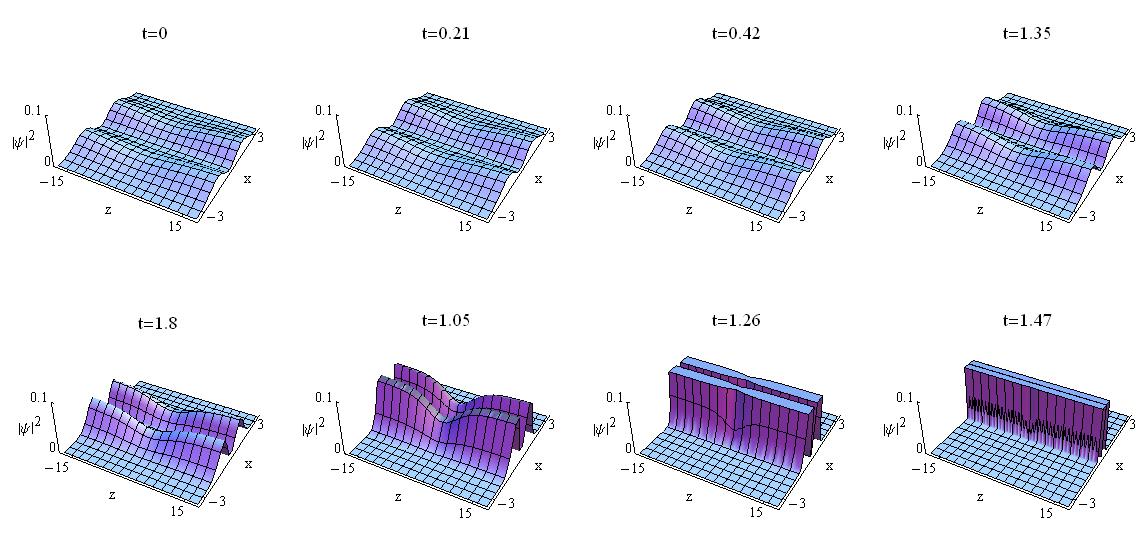}
\caption{The lowest order ($k=1$), 2D controlled, unstable BEC
state that in the parallel direction behaves as a gray soliton with
$\epsilon = 0.7$. Other parameters are the same as in Figs.
\ref{unstable_bright_k=0}-\ref{unstable_gray_k=0}.}\label{unstable_gray_k=1}
\end{figure}

The temporal evolution of the perpendicular and parallel scales,
$\sigma_x^{-1}$ and $q_{1D}(s)$, are displayed in Fig.
\ref{fig30}, together with the coefficient of the perpendicular
restoring force, $\omega_x^2(s)$. The collapse of the 2D
structures, that in the parallel direction correspond to a bright,
dark, and gray soliton is displayed in Figs.
\ref{unstable_bright_k=0}-\ref{unstable_gray_k=0}, in the case of a
ground Hermite-Gauss mode, $k=0$. The solution features an explosive
collapse within a limited period of time. The evolution of the
unstable (collapsing), first order ($k=1$) mode, is displayed in Figs.
\ref{unstable_bright_k=1}-\ref{unstable_gray_k=1}. Its behavior is
similar to that of the ground mode.

If there is an initial displacement in the parallel direction,
$R_0\ne 0$, in the course of its collapse, the soliton-like
structure will approach the equilibrium position $z=0$. The collapse
occurs roughly at the same rate in both directions, and that it
needs to be supported by a rapid (explosive) growth of the restoring
force.

The collapse of the controlled BEC structures, that occurs in the
regime $\omega_z = {\rm constant}$ and with an arbitrary dependence
$F_1(\omega_x(, \omega_y, s)=0$, is essentially 2D. It is obvious
from Eqs. (\ref{16a}) and (\ref{q1D-bis}) that the singularities of
the effective parallel and perpendicular wavenumbers scale as
$q_{1D}(s)\propto 1/[\sigma_x(s)\sigma_y(s)]\propto 1/(s-s_C)$.
Thus, if we adopt $\omega_y={\rm constant}$, we obtain a regular
expression for $\sigma_y$, Eq. (\ref{sigmadva}), i.e. the collapse
would occur in the $x,y$ plane, producing a thin 1D string along
$y$. Likewise, if we adopt the dependence $\sigma_x=\sigma_y$ (which
is a stronger requirement than $\omega_x=\omega_y$), the collapse in
the perpendicular, $x$ and $y$, directions will be the same, and
much slower than in the parallel direction. In other words, the
controlled BEC would collapse first into a pancake structure in the
perpendicular plane, that would at a later stage and on a slower
time scale, collapse also in the radial direction.

\section{Conclusions}\label{conclusion}

In this paper, we have used a newly developed analytical procedure to
construct a wide class of localized solutions of the controlled 3D
GPE governing the dynamics of BECs in the presence of a
spatio-temporally varying external potential, where the latter is
composed by a 3D parabolic time-varying potential trap plus a
controlling potential to be determined self-consistently. This has
been done on the basis of recent investigations \cite{GPE1,pre2009}.
According to these results, we have found a class of solutions in
the factorized form 
$\psi (\mathbf{r}, s) = \psi_\bot (\mathbf{r}_\bot, s) \psi_z (z,s)$, which allows us to decompose the
3D GPE into a pair of coupled partial differential equations (i.e. a
2D linear Schr\"{o}dinger equation, governing the evolution of
$\psi_\bot$ and a 1D NLSE, governing the evolution of $\psi_z$),
plus a constrained variational conditions which, among all possible
choices of $V_{contr}$, does not change the mean energy of the
system and therefore minimizes the effects introduced by our control
operation. We have extended our previous investigation \cite{pre2009}
to a wider family of localized 3D solutions of the controlled GPE. 
In particular, we have found a wide class of localized solutions whose
transverse profile is expressed in terms of breathing Hermite-Gauss
functions and the longitudinal one expressed in terms of breathing
bright, dark or grey solitons. Furthermore, we have studied in details 
the properties of the controlled BEC
states in the 2D case. It is demonstrated that they feature the
breathing due to the oscillations of the perpendicular solution, as well as the oscillations in the parallel direction, arising from the initial
displacement of the structure from the bottom of the potential well
in the parallel direction. In the examples displayed in our Figs.
\ref{fig20}--\ref{unstable_gray_k=1}, the frequency of the latter
oscillations was smaller than that of the "breathing", and the
parallel oscillations were not noticeable on the "breathing" time
scale.

The stability of the controlled structures, both bright and 
dark/gray, is governed by the shape and the temporal dependence of the
potential trap. We note that the minimization condition for the
control, Eq. (\ref{q1D-bis}), introduces an additional constraint
for the possible experimental realization of the trap. The parallel
and perpendicular coefficients of the restoring force are related,
and they can not be both adopted arbitrarily. We have demonstrated
that a stable solution can be obtained if the perpendicular
restoring force is stationary, while the parallel force periodically
changes the sign. In other words, the external parallel force needs
to be switched periodically from the confining to deconfining, in
order to arrest the inherent collapse of the soliton due to the
nonlinear interactions. Conversely, when the coefficient of the
parallel restoring force is stationary, the controlled solution is
unstable and undergoes a collapse within the finite period of time.
To maintain the control throughout the collapse, the corresponding
perpendicular restoring force needs to have a simultaneous
singularity in time. In a 3D geometry, the collapse remains
predominantly 2D. Thus, in the regime when one of the perpendicular
coefficients is constant, the controlled BEC collapses into a 1D
string, while in a symmetric situation in the perpendicular plane,
it collapses first into a pancake-like structure in the perpendicular
plane, that will collapse also in the radial direction, at a later
stage and on a slower time scale.

We want to point out that this stability analysis is not exhaustive,
but a more general stability analysis will be given in a future
work.

\end{document}